\documentclass[%
 aip,
 amsmath,amssymb,
preprint,%
]{revtex4-1}

\usepackage{graphicx}
\usepackage{dcolumn}
\usepackage{bm}

\usepackage[utf8]{inputenc}
\usepackage[T1]{fontenc}
\usepackage{mathptmx}

\begin{document}

\preprint{AIP/xxx}

\title{Acoustic response of a laser-excited polycrystalline Au-film studied by ultrafast Debye-Scherrer diffraction at a table-top short-pulse X-ray source}

\author{W. Lu}
 \affiliation{Faculty of Physics and Centre for Nanointegration Duisburg-Essen, University of Duisburg-Essen, Lotharstrasse 1, 47048 Duisburg, Germany}
 \affiliation{now at: European X-Ray Free-Electron Laser Facility, Holzkoppel 4, 22869 Schenefeld, Germany.}
\author{M. Nicoul}
\author{U. Shymanovich}
\author{F. Brinks}
\author{M. Afshari}
\author{A. Tarasevitch}
\author{D. von der Linde}
\author{K. Sokolowski-Tinten}%
\email{klaus.sokolowski-tinten@uni-due.de.}
\affiliation{Faculty of Physics and Centre for Nanointegration Duisburg-Essen, University of Duisburg-Essen, Lotharstrasse 1, 47048 Duisburg, Germany}%

\date{\today}

\begin{abstract}
The transient acoustic response of a free-standing, polycrystalline thin Au film upon femtosecond optical excitation has been studied by time-resolved Debye-Scherrer X-ray diffraction using ultrashort Cu K$_{\alpha}$ X-ray pulses from a laser-driven plasma X-ray source. The temporal strain evolution has been determined from the transient shifts of multiple Bragg diffraction peaks. The experimental data are in good agreement with the results of calculations based on the two-temperature model and an acoustic model assuming uni-axial strain propagation in the laser-excited thin film.
\end{abstract}

\maketitle

\section{\label{intro}Introduction}

Due to the rapid progress in the development of short pulse X-ray and electron sources the field of ultrafast structural dynamics has seen dramatic progress in the last two decades (e.g.\  refs.\ \citenum{schoenlein19,sciaini19} and references therein). With respect to X-rays large-scale facilities as X-ray free electron lasers (XFELs) define the current state-of-the-art\cite{bostedt16,dunne18,chapman19}, but laser-plasma based X-ray sources still represent an interesting alternative due to their simplicity, versatility, low cost and accessibility. In fact, these table-top, lab-scale sources have enabled the first ultrafast (i.e.\ femtosecond) time-resolved X-ray diffraction experiments (e.g.\ refs.\ \citenum{rischel97,rosepetruck99,rousse01,sokolowskitinten03}) and are still the subject of intense developments to improve their efficiency
\cite{weisshaupt14} or to reach higher photon energies\cite{fourmaux16,azamoum18a,azamoum18}.

Due to their comparably much lower X-ray flux these sources have been mainly used to study single-crystalline materials which provide strong diffraction signals in a reflective Bragg-geometry. However, laser-pump X-ray probe experiments often require thin film samples to match the thickness of the optically excited layer to the X-ray probing depth. Since not all materials can be grown as high quality single crystals in thin film form the range of materials, that can be studied with this approach, is limited.

In contrast, for most materials polycrystalline films can be prepared much easier. In this case the random orientation of crystallites in the sample allows to use the well-known Debye-Scherrer scheme\cite{debye16}. While the scattering signal is distributed over a diffraction ring (instead of localized diffraction spots) this scheme represents a much simpler approach since no precise sample adjustments (i.e.\ Bragg-angle) are necessary and the signal of several Bragg-peaks can be recorded simultaneously. Debye-Scherrer diffraction is easy to realize at sources that provide high X-ray flux as well as a collimated beam like synchrotrons or XFELs, enabling even single-pulse detection of diffraction patterns with high signal-to-noise (e.g.\ refs.\ \citenum{wittenberg14, gleason15, zalden19}).

The situation is much more challenging for laser-plasma based X-ray sources not only due to their lower X-ray flux, but also because of their spatially incoherent, full-solid-angle emission. Therefore, use of an appropriate X-ray optic, which collects and (quasi-)collimates the radiation is mandatory to enable this approach \cite{bargheer05, shymanovich08, shymanovich09, zamponi10,rathore17, schollmeier18}. Successful application of ultrafast Debye-Scherrer diffraction at a laser-plasma based X-ray source has been recently demonstrated by measuring transient changes of electron density for {\it thick} (10 -100 $\mu$m) powder samples of ionic crystals\cite{zamponi10, zamponi12, hauf19}.

Here we present the application of time-resolved Debye-Scherrer diffraction at a femtosecond laser-plasma based X-ray source for the study of structural dynamics in a {\it thin} sample, namely the acoustic response of a 200 nm, polycrystalline Au-film upon ultrafast optical excitation. From the measured changes of the Debye-Scherrer diffraction patterns, in particular the time-dependent shift of multiple Bragg-peaks, we determine the transient strain evolution in the sample. The measured data are in good agreement with calculations using the two-temperature-model to estimate the time-dependent pressure/stress and an acoustic model assuming uni-axial strain wave propagation in the film.

\section{\label{exp}Experimental scheme and data}

\begin{figure}[hbt]
\includegraphics[width=1\linewidth]{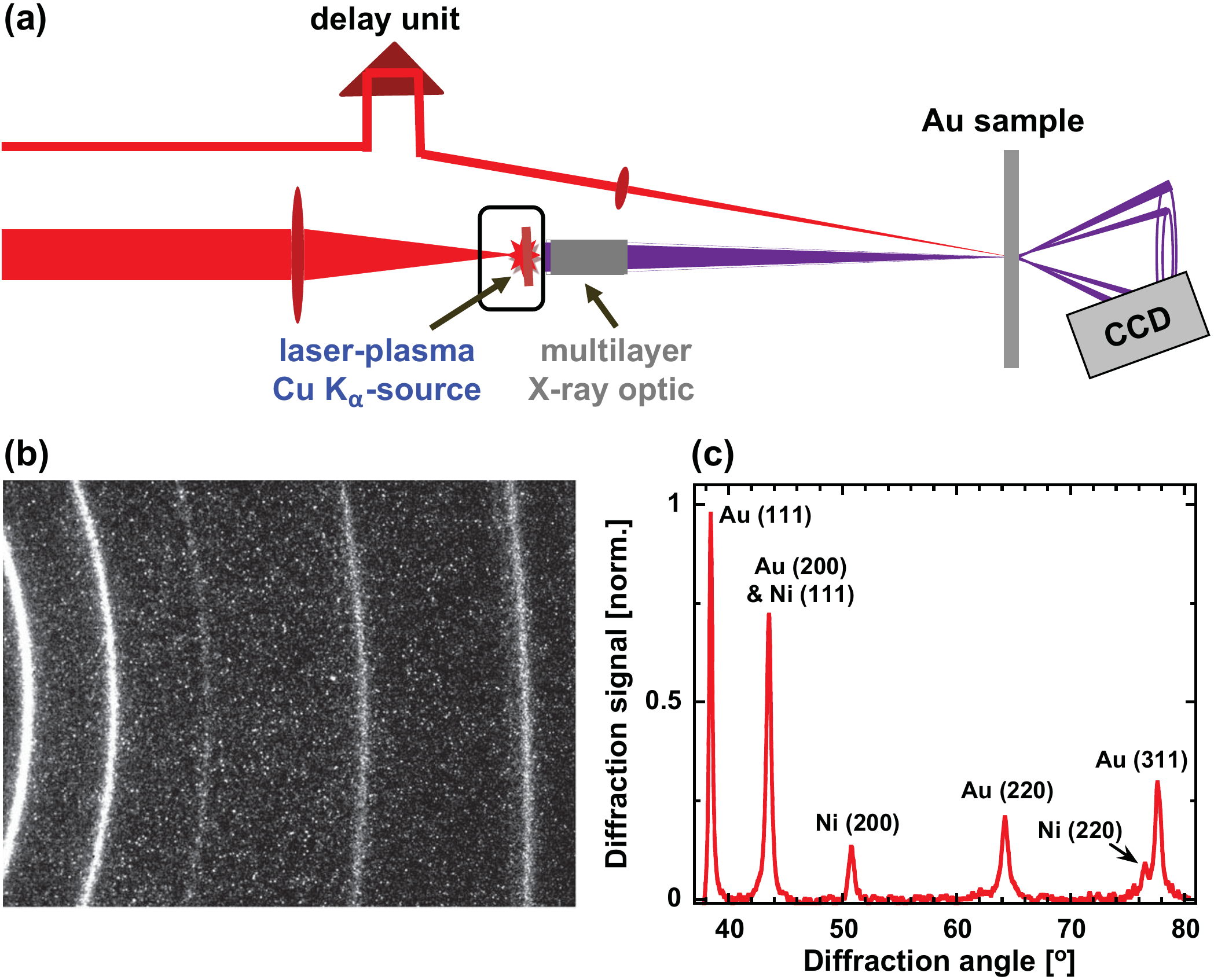}
\caption{(a): Scheme of the experimental setup. (b) Typical diffraction pattern of the 200 nm polycrystalline Au film recorded by the X-ray area detector. (c) Diffraction profile $I(\theta )$ obtained by azimuthal integration of the diffraction pattern in (b).\label{fig1}}
\end{figure}

The experiments were performed using a table-top laser-plasma based Cu K$_{\alpha}$ X-ray source and the principle scheme of the experimental setup is shown in Fig.\ \ref{fig1}(a). Short bursts of Cu K$_{\alpha}$ radiation at 8 keV were produced by focusing femtosecond laser pulses (repetition rate 10 Hz, pulse energy 150 mJ, pulse duration 120 fs, wavelength 800 nm) onto the surface of a moving 10 $\mu$m thick copper tape housed in a small vacuum chamber. A pre-pulse scheme is employed to optimize X-ray production \cite{bastiani97,lu09}, resulting in a total Cu K$_{\alpha}$-flux of more than $10^{10}$ photons per pulse.

Since the X-rays are emitted into the full solid angle we used a graded multi-layer Montel-type X-ray mirror\cite{osmic} to collect the emitted K$_{\alpha}$ radiation from the backside of the tape-target and to image the source onto the sample under study. With a magnification of 5$\times$ more than $10^5$ K$_{\alpha}$ photons per pulse are delivered to the sample in a quasi-collimated beam of 0.23$^{\circ}$ con-/divergence and a spot size of about 140 $\mu$m. The sample - a 200 nm free standing polycrystalline Au film supported by a Ni-mesh\cite{lebow} - is optically excited by a small fraction split off from the main laser beam. With a laser spot diameter (FWHM) of 400 $\mu$m on the sample, which is approximately 3$\times$ larger than the X-ray probe beam, the measured X-ray signals represent the response of a homogeneously excited region. Transient diffraction patterns, typically accumulated over 3000 pulses (5 min.\ integration time), were recorded with a single-photon sensitive phosphor-based X-ray area detector (Photonic Science X-Ray Gemstar HS) as a function of time delay between the optical pump pulse (peak fluence $\approx$ 160 mJ/cm$^2$) and the X-ray probe. In order to simultaneously record as many diffraction orders as possible, the detector was placed close to the sample (distance 38 mm) and not normal to the direct X-ray beam, but at a shallow angle of 28$^{\circ}$. This allowed us to cover an angular range of $35^{\circ} \leq \theta \leq 82^{\circ}$ ($\theta$: diffraction angle measured with respect to the incoming X-ray beam; see also Fig.\ \ref{fig3}(a)).

A typical Debye-Scherrer diffraction pattern of the Au-film, as recorded by the X-ray area detector, is depicted in Fig.\ \ref{fig1}(b). The corresponding diffraction profile $I(\theta )$, obtained by azimuthal integration, is presented in Fig.\ \ref{fig1}(c). The four lowest order diffraction peaks of Au can be clearly identified, as well as some (weaker) diffraction peaks of the supporting Ni-mesh (all diffraction peaks are labelled by their Miller indices in Fig.\ \ref{fig1}c).

\begin{figure}[hbt]
\includegraphics[width=1\linewidth]{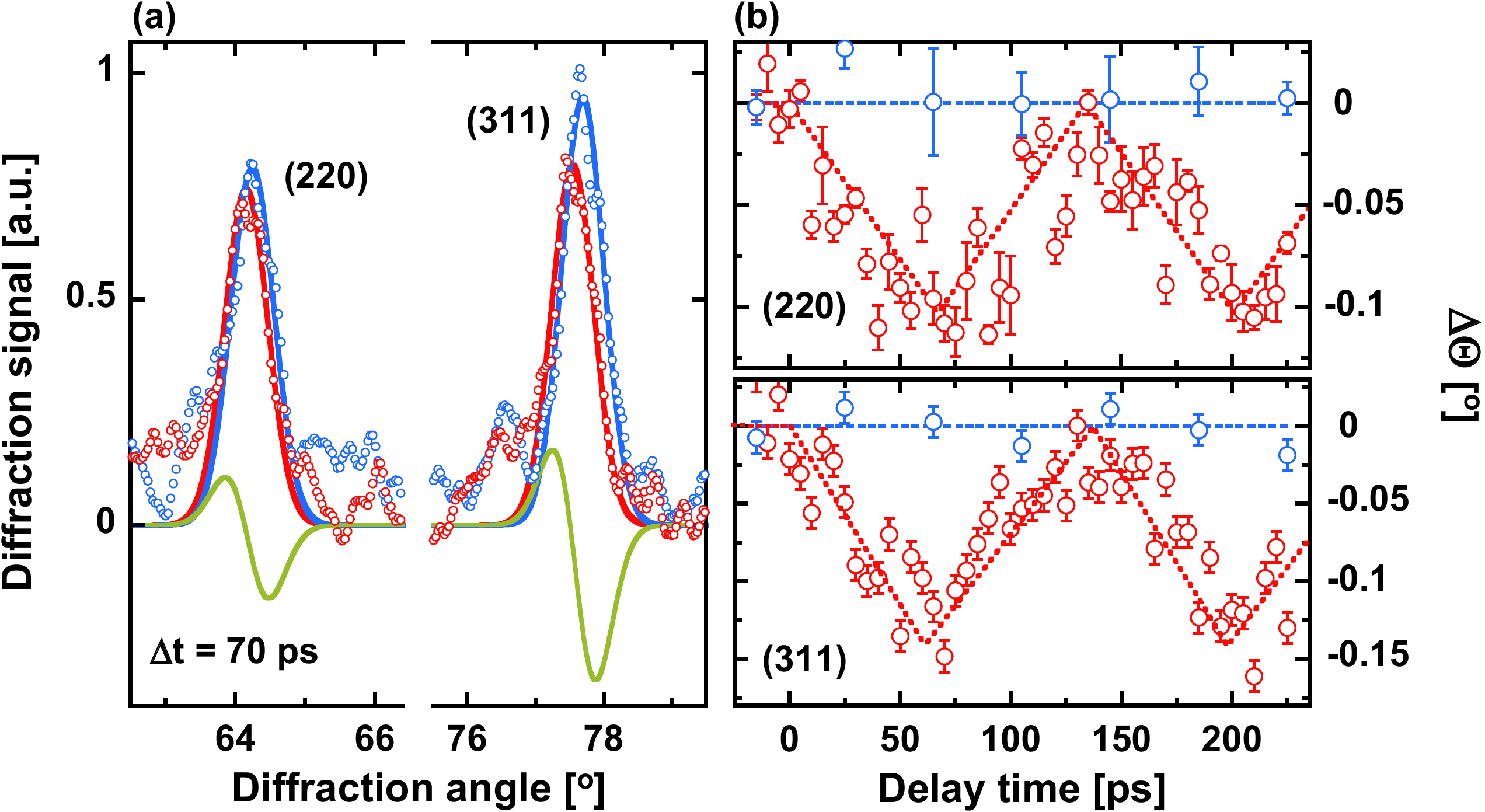}
\caption{(a): Diffraction profiles $I(\theta )$ of the (220)- (left) and (311)-reflection (right) without (blue) and with (red) laser pumping (pump-probe time delay $\Delta$t = 70 ps). Open circles: experimental data; solid lines: Gaussian fits; green solid curve: difference of the fitting results with (red curve) and without (blue curve) pumping. (b) Angular shift of the (220)- (top) and the (311)-reflection (bottom) as a function of pump-probe time delay obtained from the Gaussian fitting of the diffraction peaks. Red data points: with pumping; blue data points: reference data without pumping measured over the course of the experiment at the given delay setting. The dashed curves represent guides to the eye.\label{fig2}}
\end{figure}

In the analysis of the time-resolved data we focused on the two higher order diffraction peaks, namely the (220)- and the (311)-reflection, because of the larger magnitude of the laser-induced changes as compared to the low order peaks. For the (111)-reflection we determined only the maximum shift (see Fig.\ \ref{fig3}(b)), while the (200)-reflection could not be properly analyzed due to overlap with the Ni (111)-reflection. Results are presented in Fig.\ \ref{fig2}(a), which shows a zoom-in of the diffraction profiles $I(\theta )$ of the (220)- and (311)-reflection without (blue data points) and with (red data points) laser-pumping at a pump-probe time delay of 70 ps.

Upon pumping both diffraction peaks shift towards smaller diffraction angles and are also slightly reduced in amplitude. To quantify the shift, the Bragg-peaks have been fitted by a Gaussian function. However, to eliminate any effects on the fitting due to the noisy background as well as the adjacent Ni (220)-peak, only data points with an intensity above 30\% of the corresponding peak maximum have been used for fitting. The results of such fits are shown as red (with pump) and blue (without pump) solid lines in Fig.\ \ref{fig2}(a). The derivative-like shape of the difference of these fits (green solid curve in Fig.\ \ref{fig2}(a)) emphasizes that the peak shift represents the main pump-induced effect\cite{DW} and the following analysis will focus on this.

Fig.\ \ref{fig2}(b) shows the angular shift (red data points) of the (220) (top) and the (311) Bragg peaks, respectively, as a function of time delay, as obtained from the fitting procedure described above. The blue data points are the results of reference measurements without pumping made over the course of the experiment at the given delay setting. The dashed curves represent guides to the eye. The temporal evolution of the angular position of both diffraction peaks exhibits a pronounced oscillatory behavior with a half-period of $(67 \pm 1)$ ps. As will be discussed in the following section, this can be attributed to strain waves propagation back and forth in the thin Au-film.

\section{\label{strain}Strain analysis and modelling}
\begin{figure}[hbt]
\includegraphics[width=1\linewidth]{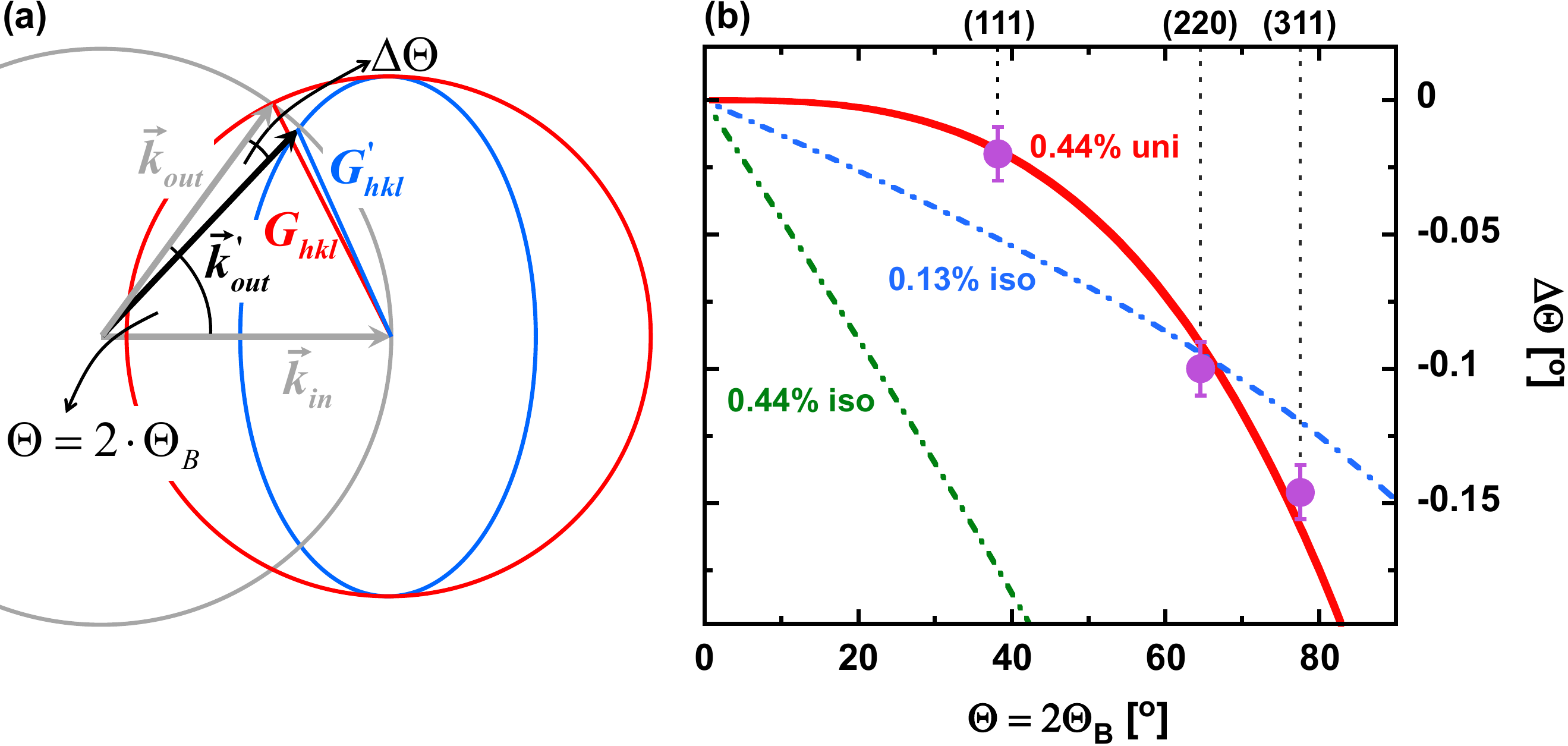}
\caption{(a) Ewald-sphere (grey circle) construction for diffraction from a polycrystalline sample without strain (red circle) and with uniaxial strain along the surface normal (blue ellipse). (b) Angular shift of the diffraction peaks as a function of diffraction angle. Violet dots: Experimental data for the (111)-, (220)-, and (311)-reflections; Green and blue dashed curves: Calculated shift assuming isotropic strain of 0.44\% (green) and 0.13\% (blue), respectively; Red solid curve: Shift calculated according to eq.\ \ref{eq1} assuming an uniaxial strain of 0.44\% along the surface normal.\label{fig3}}
\end{figure}

A shift of the diffraction peaks towards smaller diffraction angles indicates lattice expansion, i.e.\ positive strain $\eta = \Delta d_{hkl}/d_{hkl}$ ($d_{hkl}$: lattice constant). This expansion is driven by a fast, laser-induced increase of pressure, which has electronic and thermal contributions\cite{wright94}, as will be outlined in more detail below. Relaxation of the excess pressure/stress occurs by one-dimensional, longitudinal strain waves propagating normal to the surface\cite{thomson86}, because the film thickness of nominal 200 nm is much smaller than the laser excited area (diameter 400 $\mu$m). To deduce the transient strain from the measured peak shift one needs to consider, that in Debye-Scherrer configuration scattering at a particular scattering angle $\theta$ occurs by a subset of all crystallites which are oriented such that the Bragg-condition is fulfilled for a particular Bragg-peak (hkl) with $\Theta = 2\Theta_{B}$. The corresponding scattering diagram is depicted in Fig.\ \ref{fig3}(a).

Herein $\vec{k}_{in}$ and $\vec{k}_{out}$ (grey) denote the wave-vector of the incoming and scattered X-rays, respectively. There end-points lie on the so-called Ewald-sphere (grey circle). For a polycrystalline sample with randomly orientied crystallites, the corresponding reciprocal lattice vectors $\vec{G}_{hkl}$ lie on a sphere with radius $G_{hkl}$ (red circle). Where this sphere cuts the Ewald-sphere, the Bragg-condition is fulfilled, leading to scattering at $\Theta = 2\Theta_{B}$.

When the film is uniaxially expanded ($\eta > 0$) along the surface normal, which equals the direction of $\vec{k}_{in}$, the sphere with radius $G_{hkl}$ is compressed in this direction into an ellipsoid (blue) with short axis $(1-\eta)\cdot G_{hkl}$. Now the Bragg-condition is fulfilled where this ellipsoid cuts the Ewald-sphere corresponding to a differently oriented subset of crystallites. Scattering, therefore, occurs at a different direction $\vec{k}^\prime_{out}$ (black) leading to an angular shift $\Delta\theta$. According to Fig.\ \ref{fig3}(b) simple geometrical considerations lead to\cite{lu13,wittenberg14}:
\begin{equation}
\Delta\Theta = -\eta\cdot\frac{(1-\cos\Theta)^2}{\sin\Theta}
\label{eq1}
\end{equation}

\begin{figure}[hbt]
\includegraphics[width=1\linewidth]{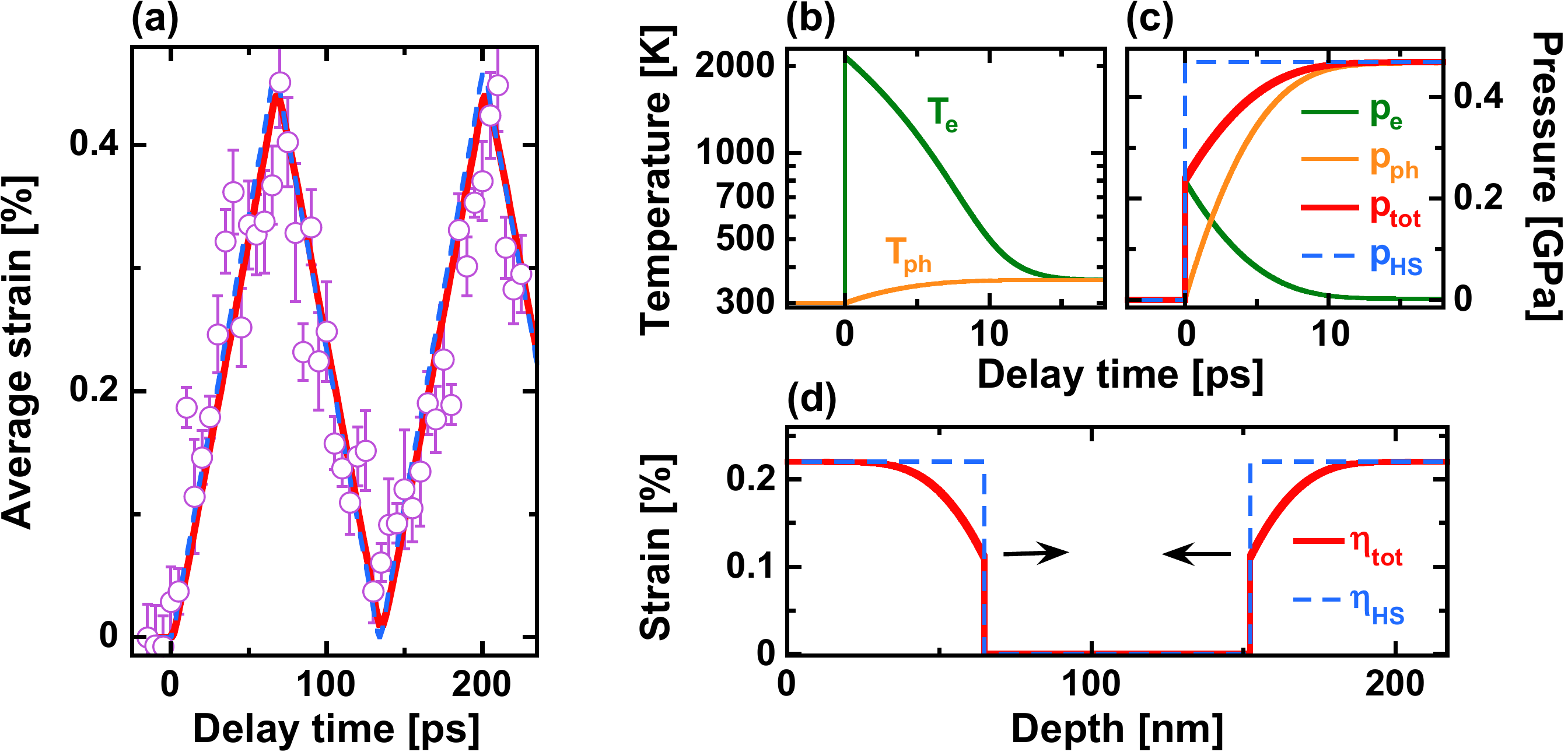}
\caption{(a) Average strain as a function of pump-probe time-delay (red open circles: experimental data; red-solid curve: calculation results (see (b) - (d)) with fully time-dependent pressure; blue-dashed curve: calculation results assuming an instantaneous increase of pressure).  (b) Electron (green) and lattice (orange) temperature as a function of time calculated using the TTM. (c) Resulting time-dependent pressure contributions (green: electronic pressure $p_e$, orange: thermal pressure $p_L$, red: total pressure $p_{tot} = p_e + p_L$). The blue-dashed curve represents a simplified instantaneous pressure increase. (d) Spatial dependence of the strain in the film 20 ps after excitation. The red solid curve is the result with the fully  time-dependent pressure as shown in (c); the blue-dashed curve corresponds to the instantaneous increase of pressure. The arrows mark the propagation direction of the strain pulses. \label{fig4}}
\end{figure}

Fig.\ \ref{fig3}(b) compares the measured maximum shift of the (111)-, (220)-, and (311)-reflection (violet data points; the (200)-peak could not be analyzed due to overlap with the strong (111)-peak of Ni) to calculations for different strain conditions. The blue and green-dashed curves represent the expected angular shifts assuming isotropic strain of 0.13\% (blue) and 0.44\% (green) ($\Delta\Theta=-\tan\Theta\cdot\eta$), respectively, which are obviously unable to describe the measured data. In contrast, good agreement is found using eq.\ \ref{eq1} with a strain of 0.44\% (red solid curve), giving clear evidence that the laser-driven expansion of the film is indeed uniaxial. Applying eq.\ \ref{eq1} also to the measured time-dependent shifts of the (220)- and (311)-peaks (compare Fig.\ \ref{fig2}(b)) the strain as a function of pump-probe time-delay can be obtained, as depicted in Fig.\ \ref{fig4}(a). Similar as the measured angular shifts the time dependent strain exhibits an (undamped - over the measured delay range) of oscillatory behavior with a half-period of $(67 \pm 1)$ ps.

As mentioned above, this oscillatory behavior is caused by strain waves travelling back and forth in the Au film. The half-period of $t_{ac}=(67 \pm 1)$ ps corresponds to the time such a strain wave needs to propagate (with the speed of sound $c_S$) through the full film thickness $d$ once. With $c_S$ = 3.24 km/s for polycrystalline Au\cite{crc_hcp05} this allows to determine the actual film thickness $d=c_S\cdot t_{ac}= (217 \pm 3)$ nm, in good agreement with the nominal film thickness and the $\pm$10\% thickness tolerance specified by the manufacturer.

To quantitatively model the response of the Au-film we applied the two-temperature-model\cite{anisimov74} (TTM) in combination with a solution of the one-dimensional elastic equations\cite{thomson86}. The relevant TTM-parameters are a constant lattice specific heat of $C_{L} = 2.5$ MJ/(m$^3$K)\cite{crc_hcp05}, an electronic specific heat $C_e(T_e)=A_e\cdot T_e$ with $A_e=67.6$ J/(m$^3$K$^2$)\cite{wright94} and an electron-phonon coupling paramter $g=1.7\times 10^{16}$ W/(m$^3$K)\cite{sokolowski17}. For simplicity we assume that laser excitation leads to an (i) instantaneous and (ii) spatially homogenous increase of the electronic temperature in the film, which is justified by (i) the relatively weak electron-phonon coupling, leading to correspondingly {\it long} electron-lattice equilibration times of a few ps\cite{sokolowski17}, and (ii) the very efficient, ballistic/superdiffusive electronic transport in Au\cite{hohlfeld97, hohlfeld00}. Results of such calculations are shown in Fig.\ \ref{fig4}(b) for an asymptotic rise in lattice temperature $\Delta T_{L,\infty} = 61$ K (see below), corresponding to an initial electron temperature $T_{e,max} = 2153$ K.

Both, the changes of the electronic and the lattice temperature cause an increase of pressure (isotropic stress) in the material, which can be expressed as\cite{wright94} $\delta P = \gamma _e C_e \delta T_e + \gamma _L C_L \delta T_L$. Herein $\gamma _e$ and $\gamma _L$ denote the electronic and lattice Gr\"uneisen parameter, respectively, with $\gamma _L = 3$\cite{barron80} and $\gamma _e/\gamma_L \approx 0.5$\cite{sokolowski17}. The resulting time-dependent electronic ($p_e$) and thermal ($p_L$) pressure contributions are depicted in Fig.\ \ref{fig4}(c) as the green and orange curve, respectively. The total pressure $p_{tot} = p_e + p_L$ (red curve) is then used to solve the one-dimensional elastic equations\cite{thomson86}. In this model the peak strain is given by:
\begin{equation}
\eta_{max} = \frac{6B\beta}{c_S^2\rho}\cdot\Delta T_{L,\infty}
\label{eq2}
\end{equation}
Herein $B = 177$ GPa denote the bulk modulus\cite{anderson89}, $\beta = 1.426\times 10^{-5}$ K$^{-1}$ the linear thermal expansion coefficient\cite{anderson89}, $\rho = 19.3$ g/cm$^3$ the density, and $c_S$ = 3.24 km/s the sound velocity\cite{crc_hcp05} of Au. Eq.\ \ref{eq2} together with the measured $\eta _{max} = 0.44$\% results in $\Delta T_{L,\infty} = 61$ K, the value that has been used in the TTM-calculations.

Results of the acoustic modelling are presented in Fig.\ \ref{fig4}(d), which shows the calculated strain profile at $\Delta t = 20$ ps. Since the experiments have been carried out on a free-standing film, strain waves with an equal amplitude of $0.5\cdot\eta _{max}$ are launched at both film surfaces, which propagate into the film. At the chosen time of $\Delta t = 20$ ps they have travelled less than 1/3 of the film thickness and are, therefore, still spatially separated. The spatial shape of each pulse reflects the temporal evolution of the total pressure $p_{tot}$. At later times both pulses overlap and when approaching the opposite (free) surface they are reflected with a sign change (wave reflection at an open end).

Since the time-dependent spatial strain distribution is inhomogeneous, the changes of the X-ray diffraction patterns are characterized by angular shifts as well as changes in shape (e.g.\ broadening) of the rocking curves of individual diffraction peaks\cite{nicoul11}. However, due to the large angular width of the Bragg-peaks of about 0.7$^{\circ}$ in our experiment such shape changes of the rocking curves can not be resolved. As a result, the measured shifts represent the average strain in the film at a given time, as presented in Fig.\ \ref{fig4}(a). The average strain, as determined from the acoustic model calculations (red solid curve in Fig.\ \ref{fig4}(a)) agrees very well with the experimental data.

Over the measured delay range the strain oscillations are undamped, as has been already emphasized above. This is a consequence of having a free-standing film as sample, where the acoustic pulses exhibit total reflection at the free surfaces. In contrast, strain waves in thin films on substrates are (partially - depending on the difference in acoustic impedance between film and substrate) transmitted into the substrate upon each round-trip and experience, therefore, damping\cite{nicoul11}.

We also performed the acoustic modelling with a simplified pressure evolution, namely an instantaneous increase at $\Delta t = 0$ (equivalent to an infinitely fast electron-phonon coupling and/or equal Gr\"uneisen parameters), as indicated by the blue-dashed curve in Fig.\ \ref{fig4}(c). The resulting strain pulses (now exhibiting a rectangular shape) and the corresponding time-dependence of the average strain are depicted by the blue-dashed curves in Figs.\ \ref{fig4}(d) and \ref{fig4}(a), respectively. With the given accuracy our current data do not allow to discriminate between the two scenarios. However, our previous experiments on an epitaxial Au-film\cite{nicoul11} provided clear evidence, that the finite electron-phonon coupling as well as the difference of the Gr\"uneisen parameters need to be taken into account to properly interpret the observed acoustic response. 

\section{\label{sum}Summary}
In summary, we have used ultrafast time-resolved Debye-Scherrer diffraction to study the acoustic response of a free standing polycrystalline Au film upon femtosecond optical excitation. From the measured shifts of different Bragg-peaks the transient strain evolution in the film has been determined, which follows the behavior expected from longitudinal acoustic waves propagating normal to the surface of the free standing sample. Very good quantitative agreement is found with the results of calculations based on the two-temperature model and a solution of the one-dimensional acoustic equations taking into account electronic and thermal contributions to the laser-induced pressure/stress. Our results demonstrate the feasibility of time-resolved Debye-Scherrer diffraction experiments on {\it thin} solid samples (where film thickness and excitation depth are matched) using a laser-plasma based X-ray source. This considerably extends the range of materials for which Debye-Scherrer diffraction at such sources can be applied, in particular when future prospects to increase their efficieny\cite{weisshaupt14} are considered.

\begin{acknowledgments}
Financial support by the {\it Deutsche Forschungsgemeinschaft} (DFG, German Research Foundation) through project C01 {\it Structural Dynamics in Impulsively Excited Nanostructures} of the Collaborative Research Center SFB 1242 {\it Non-Equilibrium Dynamics of Condensed Matter in the Time Domain} (project number 278162697) is gratefully acknowledged.
\end{acknowledgments}


\begin{thebibliography}{43}
\bibitem [{DW()}]{DW}%
  \BibitemOpen
  \href@noop {} {}\bibinfo {note} {The limited accuracy of the experiment did
  not allow to determine the reduction of the Bragg-peak intensities due to the
  Debye-Waller effect. In fact, using published temperature-dependent
  Debye-Waller factors for Au\cite{synecek70,gao99} the reduction of the peak
  intensity expected for a temperature rise of 61 K is only 4\% for the
  (311)-reflection, the highest order reflection we did measure.}\BibitemShut
  {Stop}%
\end{thebibliography}
%

\end{document}